\documentclass{article}
\usepackage{graphicx}
\usepackage[english]{babel}

\title{
\includegraphics[width=0.35\textwidth]{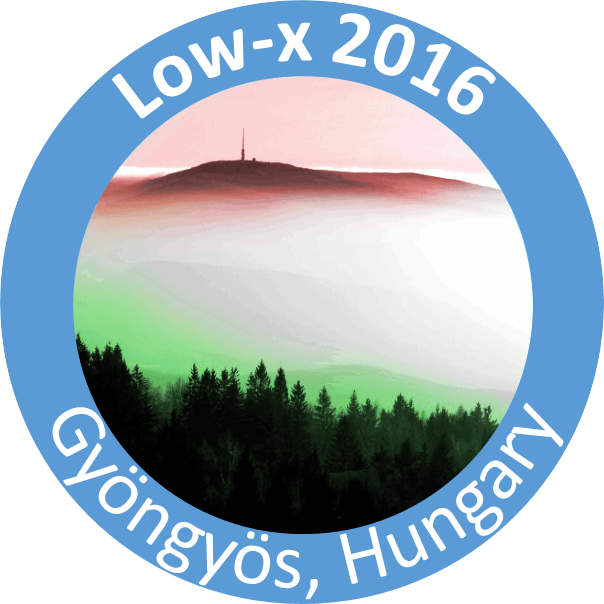}\\[1cm]
Exclusive production of two and four pions \\in proton-proton scattering}
\author{{P. Lebiedowicz$^1$, O. Nachtmann$^2$, A. Szczurek$^1$
\footnote{Also at University of Rzesz\'ow, PL-35959 Rzesz{\'o}w, Poland.}}\\[1ex]
$^1$Institute of Nuclear Physics Polish Academy of Sciences,\\Radzikowskiego 152, PL-31342 Krak{\'o}w, Poland\\
$^2$Institut f\"ur Theoretische Physik, Universit\"at Heidelberg,\\Philosophenweg 16, D-69120 Heidelberg, Germany\\
}

\begin{document}

\fontfamily{lmss}\selectfont
\maketitle

\begin{abstract}
We consider exclusive $pp \to pp \pi^{+} \pi^{-}$ 
and $pp \to pp \pi^{+} \pi^{-}\pi^{+} \pi^{-}$ 
reactions at high energies.
The calculation is based on a tensor pomeron model
and the amplitudes for the processes are formulated 
in an effective field-theoretic approach.
In the case of $pp \to pp \pi^{+} \pi^{-}$ process we consider
both diffractive and photoproduction mechanisms
and we include the non-resonant $\pi^{+}\pi^{-}$ continuum and the resonance 
$f_{0}(500)$, $f_{0}(980)$, $f_{2}(1270)$, $\rho(770)$ contributions.
We discuss how two pomerons couple to tensor meson $f_{2}(1270)$
and the interference effects of resonance and dipion continuum.
We find that the relative contribution of resonances $\rho(770)$, $f_2(1270)$ and
dipion continuum strongly depends on the cut on proton transverse momenta. 
In the case of exclusive central $4 \pi$ production
we include the contribution via the intermediate 
$\sigma\sigma$ and $\rho\rho$ states.
For both processes the theoretical results have been 
compared with the experimental data
and predictions for planned or being carried out experiments 
(e.g. STAR, ATLAS-ALFA) are presented.
\end{abstract}

\section{Introduction}
\label{intro}
Central production mediated by the ``fusion'' of two exchanged pomerons
\cite{Lebiedowicz:2016ioh,Fiore:2015lnz}
is an important diffractive process for the investigation of properties of dipion resonances,
in particular, for search of gluonic bound states (glueballs).
The experimental groups at the CERN-ISR \cite{Breakstone:1986xd},
COMPASS \cite{Austregesilo:2016sss}, 
STAR \cite{Adamczyk:2014ofa}, 
CDF \cite{Aaltonen:2015uva}, 
ALICE \cite{Schicker:2012nn}, and
CMS \cite{CMS:2015diy}
all show visible structures in the $\pi^{+}\pi^{-}$ invariant mass.
The LHCb experiment is also well suited to measuring central
exclusive production processes \cite{McNulty:2016sor}.

Some time ago two of us have formulated a Regge-type model 
of the dipion continuum
for the exclusive reaction $pp \to pp \pi^{+}\pi^{-}$ with parameters 
fixed from phenomenological analysis of total and elastic $NN$ and $\pi N$ 
scattering \cite{Lebiedowicz:2009pj}.
The model was extended to include rescattering corrections 
due to $pp$ nonperturbative interaction \cite{Lebiedowicz:2011nb,Staszewski:2011bg}.
The exclusive reaction $pp \to pp \pi^{+}\pi^{-}$
constitutes an irreducible background 
to the scalar $f_{0}(1500)$ \cite{Szczurek:2009yk} and $\chi_{c0}$ \cite{Lebiedowicz:2011nb} mesons production.
These model studies were extended to the exclusive $pp \to ppK^{+}K^{-}$ reaction \cite{Lebiedowicz:2011tp}.
The largest uncertainties in the model are due
to the unknown off-shell pion form factor and the absorption effects; see Ref.~\cite{Lebiedowicz:2015eka}.
Such an approach gives correct order of magnitude cross sections,
however, does not include resonance contributions which interfere with the continuum contribution.

First calculations of central exclusive diffractive production of $\pi^{+} \pi^{-}$ continuum 
together with the dominant scalar $f_{0}(500)$, $f_{0}(980)$, 
and tensor $f_{2}(1270)$ resonances was performed in Ref.~\cite{Lebiedowicz:2016ioh}.
Here we use the tensor-pomeron model formulated in \cite{Ewerz:2013kda}; 
see also \cite{Nachtmann:1991ua}.
In this model pomeron exchange is effectively treated as the exchange of a rank-2 symmetric tensor.
In \cite{Ewerz:2016onn} we show that the tensor pomeron is consistent 
with the STAR experimental data on polarised high-energy $pp$ elastic scattering~\cite{Adamczyk:2012kn}.
In Ref.~\cite{Lebiedowicz:2013ika} the model was applied to 
the diffractive production of several scalar and pseudoscalar mesons in the reaction $p p \to p p M$.
The corresponding pomeron-pomeron-meson coupling constants are not known 
and have been fitted to existing WA102 experimental data.
In most cases one has to add coherently amplitudes 
for two pomeron-pomeron-meson couplings 
with different orbital angular momentum and spin of two ``pomeron particles''.
\footnote{We wish to emphasize that the tensorial pomeron can
equally well describe the WA102 experimental data 
on the exclusive meson production as the less 
theoretically justified vectorial pomeron frequently used in the literature. 
The existing low-energy experimental data do not allow to
clearly distinguish between the two approaches as the presence of subleading
reggeon exchanges is at low energies very probable for many $p p \to p p M$ reactions.}
In \cite{Bolz:2014mya} an extensive study of the photoproduction reaction
$\gamma p \to \pi^{+} \pi^{-} p$ was presented.
The resonant ($\rho^0 \to \pi^{+}\pi^{-}$) and non-resonant (Drell-S\"oding)
photon-pomeron/reggeon $\pi^{+} \pi^{-}$ production in $pp$ collisions 
was studied in \cite{Lebiedowicz:2014bea}. 

The identification of glueballs can be very difficult.
The studies of different decay channels in central
exclusive production would be very valuable in this context.
One of the possibilities is the $p p \to p p \pi^+ \pi^- \pi^+ \pi^-$
reaction being analysed at the RHIC and LHC.
In Ref.~\cite{Lebiedowicz:2016zka} we analysed
the exclusive diffractive production of four-pion
via the intermediate $\sigma \sigma$ and $\rho \rho$ states
within the tensor-pomeron model.

\section{Sketch of the formalism}
\label{formalism}

\begin{figure}[ht]
\centering
\includegraphics[width=3.6cm,clip]{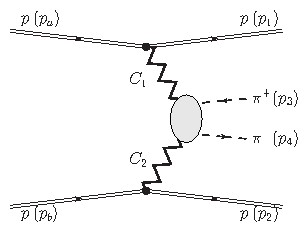}
\includegraphics[width=4.5cm,clip]{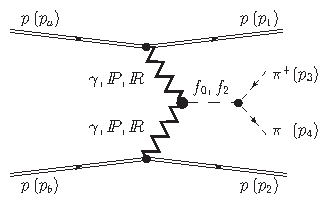}
\includegraphics[width=3.8cm,clip]{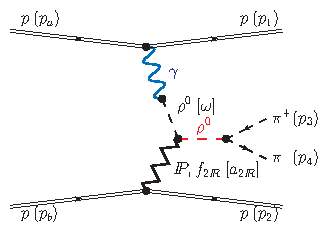}
\caption{
Generic Born-level diagrams for central exclusive production of continuum $\pi^+ \pi^-$
and resonances in proton-(anti)proton collisions.
Here we labelled the exchanged objects by their charge conjugation numbers
$C_{1}$, $C_{2} \in \lbrace +1, -1 \rbrace$.
}
\label{fig:0}
\end{figure}

The Born-level diagrams for the continuum and resonant $\pi^{+} \pi^{-}$ production
are shown in Fig.~\ref{fig:0}.
The purely diffractive amplitude is a sum of continuum amplitude 
and the amplitudes with the $s$-channel scalar and tensor resonances:
%
\begin{eqnarray} \label{amplitude_pomTpomT}
{\cal M}_{pp \to pp \pi^{+} \pi^{-}} =
{\cal M}^{\pi \pi{\rm-continuum}}_{pp \to pp \pi^{+} \pi^{-}} + 
{\cal M}^{(I\!\!P I\!\!P \to f_{0} \to \pi^{+}\pi^{-})}_{\lambda_{a} \lambda_{b} \to \lambda_{1} \lambda_{2} \pi^{+}\pi^{-}}
 + {\cal M}^{(I\!\!P I\!\!P \to f_{2} \to \pi^{+}\pi^{-})}_{\lambda_{a} \lambda_{b} \to \lambda_{1} \lambda_{2} \pi^{+}\pi^{-}}\,.
\end{eqnarray}
The Born amplitude, for instance, for the process
$pp \to pp (f_{2} \to \pi^{+}\pi^{-})$ can be written 
in the effective tensor pomeron approach as
\begin{eqnarray}
&{\cal M}^{(I\!\!P I\!\!P \to f_{2}\to \pi^{+}\pi^{-})}_{\lambda_{a} \lambda_{b} \to \lambda_{1} \lambda_{2} \pi^{+}\pi^{-}} 
=  (-i)\,
\bar{u}(p_{1}, \lambda_{1}) 
i\Gamma^{(I\!\!P pp)}_{\mu_{1} \nu_{1}}(p_{1},p_{a}) 
u(p_{a}, \lambda_{a})\,
i\Delta^{(I\!\!P)\, \mu_{1} \nu_{1}, \alpha_{1} \beta_{1}}(s_{1},t_{1}) \nonumber \\
& \qquad \qquad \qquad \quad \times 
i\Gamma^{(I\!\!P I\!\!P f_{2})}_{\alpha_{1} \beta_{1},\alpha_{2} \beta_{2}, \rho \sigma}(q_{1},q_{2}) \;
i\Delta^{(f_{2})\,\rho \sigma, \alpha \beta}(p_{34})\,
i\Gamma^{(f_{2} \pi \pi)}_{\alpha \beta}(p_{3},p_{4}) \nonumber \\
& \qquad \qquad \qquad \qquad \times 
i\Delta^{(I\!\!P)\, \alpha_{2} \beta_{2}, \mu_{2} \nu_{2}}(s_{2},t_{2}) \;
\bar{u}(p_{2}, \lambda_{2}) 
i\Gamma^{(I\!\!P pp)}_{\mu_{2} \nu_{2}}(p_{2},p_{b}) 
u(p_{b}, \lambda_{b}) \,,
\label{amplitude_f2_pomTpomT}
\end{eqnarray}
where $t_{1} = q_{1}^{2} = (p_{1} - p_{a})^{2}$, 
$t_{2} = q_{2}^{2} = (p_{2} - p_{b})^{2}$, 
$s_{1} = (p_{a} + q_{2})^{2} = (p_{1} + p_{34})^{2}$,
$s_{2} = (p_{b} + q_{1})^{2} = (p_{2} + p_{34})^{2}$,
$p_{34} = p_{3} + p_{4}$. 
$\Delta^{(I\!\!P)}$ and $\Gamma^{(I\!\!P pp)}$ 
denote the effective pomeron propagator and proton vertex function, respectively.
For the explicit expressions, see Sec.~3 of \cite{Ewerz:2013kda}.
In Ref.~\cite{Lebiedowicz:2016ioh} (see Appendix~A) 
we have considered all possible tensorial structures 
for the $I\!\!P I\!\!P f_{2}$ coupling.
%
%
For a more details, as form of form factors, 
the tensor-meson propagator $\Delta^{(f_{2})}$ and 
the $f_{2} \pi \pi$ vertex, see Refs.~\cite{Ewerz:2013kda,Lebiedowicz:2016ioh}.

We consider also the production of $\rho(770)$ resonance
and the non-resonant (Drell-S\"oding) $\pi^{+} \pi^{-}$ continuum
produced by photon-pomeron and photon-$f_{2 I\!\!R}$ mechanisms 
studied in detail in \cite{Lebiedowicz:2014bea}.
The $I\!\!P \rho \rho$ vertex is given in \cite{Ewerz:2013kda} by formula (3.47).
The coupling parameters of Regge exchanges was fixed based on
the HERA experimental data for the $\gamma p \to \rho^{0} p$ reaction.
In \cite{Lebiedowicz:2014bea} we showed that the $\rho^{0}$ term
interfere with the non-resonant terms 
producing a skewing of the $\rho^{0}$-meson line shape.
Due to the photon propagators occurring in diagrams
we expect these processes to be most important when at least one of the protons
undergoes only a very small $|t_{1,2}|$.
\section{Selected results}
\label{results}

\begin{figure}
\centering
\includegraphics[width=5.5cm,clip]{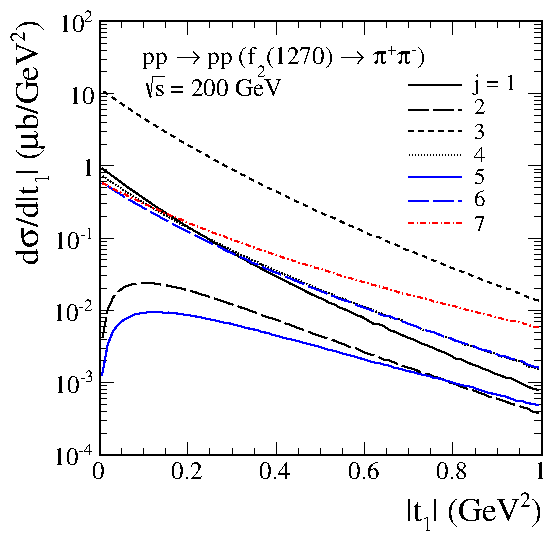}
\includegraphics[width=5.5cm,clip]{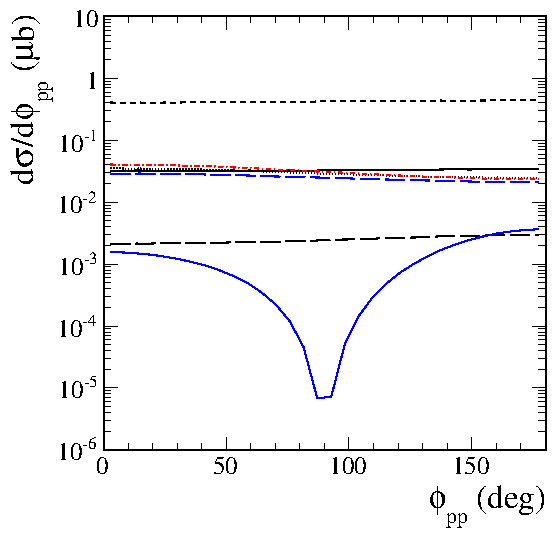}
  \caption{\label{fig:dsig_dt1}
  \small
The (Born-level) distribution in transferred four-momentum squared
between the initial and final protons (left panel) and
the distribution in azimuthal angle between the outgoing protons (right panel)
at $\sqrt{s} = 200$~GeV and $|\eta_{\pi}|< 1$.
We show the individual contributions of the different pomeron-pomeron-$f_{2}(1270)$ couplings.
For illustration the results have been obtained 
with coupling constants fixed at  $g^{(j)}_{I\!\!P I\!\!P f_{2}} = 1.0$.
}
\end{figure}

We start from a discussion of some dependences
for the central exclusive production of the $f_{2}(1270)$ meson.
For a detailed study of $f_{2}(1270)$ production see Ref.~\cite{Lebiedowicz:2016ioh}.
In Fig.~\ref{fig:dsig_dt1} we present results for 
individual pomeron-pomeron-$f_{2}$ coupling terms
(there are 7 possible terms \cite{Lebiedowicz:2016ioh})
at $\sqrt{s} = 200$~GeV and $|\eta_{\pi}|< 1$.
The different predictions differ considerably which could be checked experimentally.
We show that only in two cases ($j=2$ and 5)
the cross section $d\sigma/d|t_{1,2}|$ vanishes when $|t_{1,2}| \to 0$.
\begin{figure}
\centering
\includegraphics[width=5.5cm,clip]{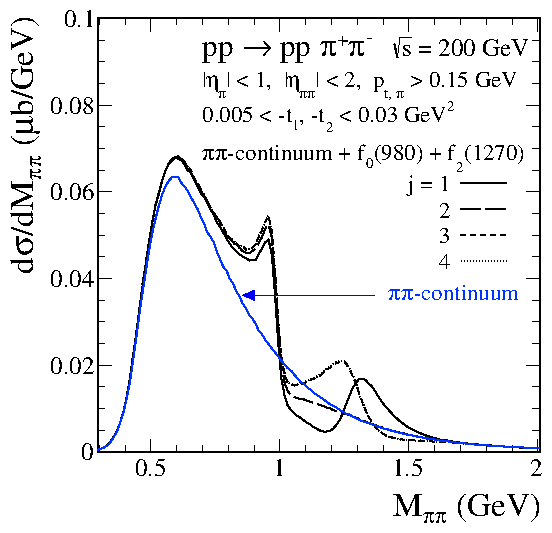}
\includegraphics[width=5.5cm,clip]{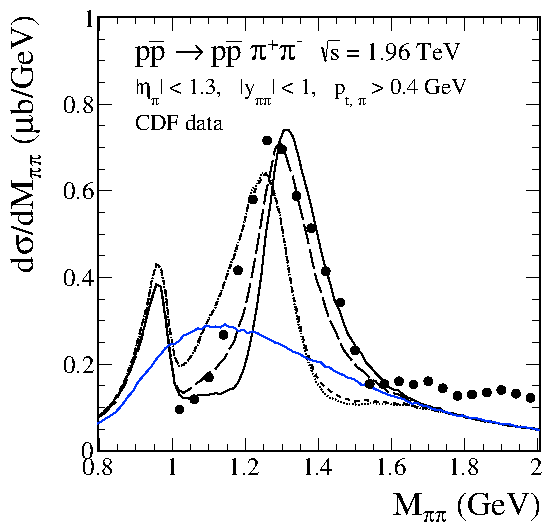}
  \caption{\label{fig:1}
  \small
Two-pion invariant mass distribution for the STAR \cite{Adamczyk:2014ofa} (left)
and CDF \cite{Aaltonen:2015uva} (right) kinematics.
The individual contributions of different $I\!\!P I\!\!P f_{2}$ couplings ($j = 1, ..., 4$)
are compared with the CDF data \cite{Aaltonen:2015uva}.
The Born calculations for $\sqrt{s}=200$~GeV 
and $\sqrt{s}=1.96$~TeV were multiplied 
by the gap survival factors $\langle S^{2}\rangle = 0.2$
and $\langle S^{2}\rangle = 0.1$, respectively.
The blue solid lines represent the non-resonant continuum contribution only 
($\Lambda_{off,M} = 0.7$~GeV) 
while the black lines represent a coherent sum of non-resonant continuum, 
$f_{0}(980)$ and $f_{2}(1270)$ resonant terms. 
}
\end{figure}
In \cite{Lebiedowicz:2016ioh}
we tried to understand whether one can approximately describe
the dipion invariant mass distribution observed by different experiments
assuming only one of the seven possible $I\!\!P I\!\!P f_{2}$ tensorial couplings. 
We found that the feature of the $\pi^{+}\pi^{-}$ distribution depends
on the cuts used in a particular experiment
(usually the $t$ cuts are different for different experiments).
As can be clearly seen from Fig.~\ref{fig:1} 
different $I\!\!P I\!\!P f_{2}$ couplings generate different interference patterns
around $M_{\pi\pi} \sim 1.27$~GeV.
A sharp drop around $M_{\pi\pi} \sim 1$~GeV is attributed to the interference
of $f_{0}(980)$ and continuum.
We can observe that the $j=2$ coupling gives results close to those observed by the CDF Collaboration \cite{Aaltonen:2015uva}.
In this preliminary study we did not try to fit the existing data \cite{Aaltonen:2015uva} 
by mixing different couplings because the CDF data are not fully exclusive
(the outgoing $p$ and $\bar{p}$ were not measured).
The calculations were done at Born level
and the absorption corrections were taken into account
by multiplying the cross section
by a common factor $\langle S^{2}\rangle$ obtained from \cite{Lebiedowicz:2015eka}.
The two-pion continuum was fixed by choosing a form factor for the off-shell pion 
$\hat{F}_{\pi}(k^{2})=\frac{\Lambda^{2}_{off,M} - m_{\pi}^{2}}{\Lambda^{2}_{off,M} - k^{2}}$ and
$\Lambda_{off,M} = 0.7$~GeV.

\begin{figure}[ht]
\centering
\includegraphics[width=5.5cm,clip]{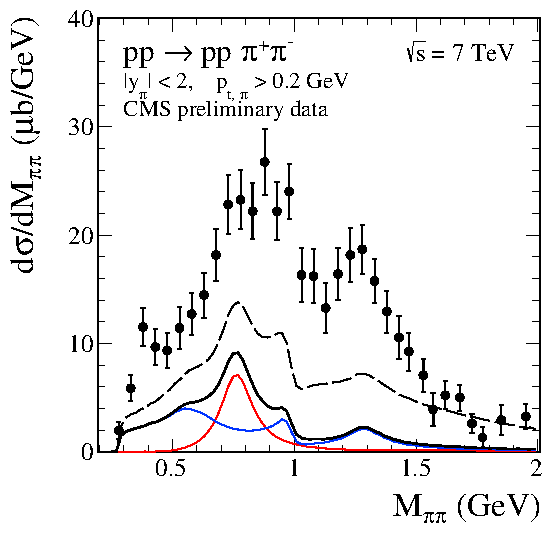}
\includegraphics[width=5.5cm,clip]{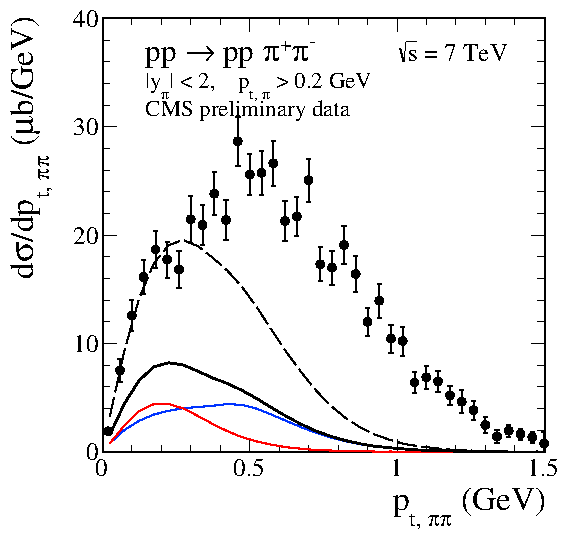}
  \caption{\label{fig:2}
  \small
The distributions for two-pion invariant mass (left panel)
and transverse momentum of the pion pair (right panel)
for the CMS kinematics at $\sqrt{s}=7$~TeV. 
Both photoproduction (red line) and purely diffractive (blue line) contributions 
multiplied by the factors $\langle S^{2}\rangle = 0.9$
and $\langle S^{2}\rangle = 0.1$, respectively, are included.
The complete results correspond to the black solid line ($\Lambda_{off,M} = 0.7$~GeV) 
and the dashed line ($\Lambda_{off,M} = 1.2$~GeV).
The CMS preliminary data scanned from \cite{CMS:2015diy} are shown for comparison.
}
\end{figure}
In Fig.~\ref{fig:2} we show results including 
in addition to the non-resonant $\pi^{+}\pi^{-}$ continuum, 
the $f_{2}(1270)$ and the $f_{0}(980)$ resonances,
the contribution from photoproduction 
($\rho^{0} \to \pi^{+}\pi^{-}$, Drell-S\"oding mechanism), 
as well as the $f_{0}(500)$ resonant contribution.
Our predictions are compared with the CMS preliminary data \cite{CMS:2015diy}.
Here the absorption effects lead to huge damping of the cross section 
for the purely diffractive term (the blue lines) 
and relatively small reduction of the cross section 
for the photoproduction term (the red lines).
Therefore we expect one could observe the photoproduction contribution.
The CMS measurement \cite{CMS:2015diy} is not fully exclusive and
the $M_{\pi\pi}$ and $p_{t,\pi\pi}$ spectra contain contributions associated
with other processes, e.g., when one or both protons undergo dissociation.
In addition, the dashed line corresponds to results with $\Lambda_{off,M} = 1.2$~GeV
and better describe the preliminary CMS data.
If we used the set of parameters adjusted to the CDF data \cite{CMS:2015diy}
for the STAR or CDF measurements our theoretical results 
there would be above the preliminary STAR data \cite{Adamczyk:2014ofa}
at $M_{\pi\pi} > 1$~GeV and in complete disagreement 
with the CDF data from \cite{Aaltonen:2015uva}.
Only purely central exclusive data expected from
CMS-TOTEM and ATLAS-ALFA will allow to draw definite conclusions.

\begin{figure}[!ht]
\centering
\includegraphics[width=5.5cm,clip]{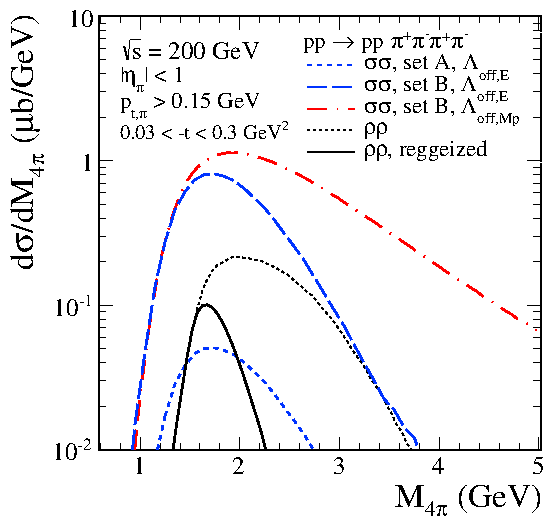}
\includegraphics[width=5.5cm,clip]{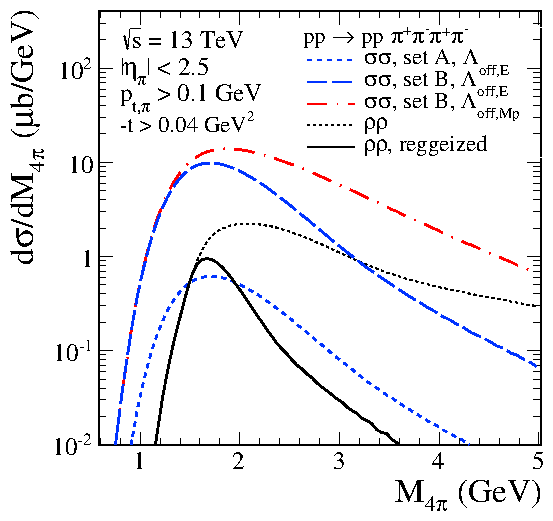}
  \caption{\label{fig:M4pi_PP}
  \small
The $4 \pi$ invariant mass distributions (for different experimental cuts)
multiplied by the factors $\langle S^{2}\rangle$ =
0.30 (for $\sqrt{s}$ = 200~GeV) and 0.23 (for $\sqrt{s}$ = 13~TeV)
estimated within the eikonal approximation (only the $pp$ rescattering).
The blue and red lines for the $\sigma \sigma$ contribution
for the exponential off-shell meson form factors ($\Lambda_{off,E} = 1.6$~GeV) 
and the monopole ones ($\Lambda_{off,M} = 1.6$~GeV), respectively.
The black lines represent results for the $\rho\rho$ contribution
without (the dotted line) and with (the solid line) 
the $\rho$ meson reggeization.
}
\end{figure}

In Fig.~\ref{fig:M4pi_PP} we show the four-pion invariant mass
distributions for the the reaction $pp \to pp \pi^{+} \pi^{-} \pi^{+} \pi^{-}$
proceeding via the intermediate $\sigma \sigma$ and the $\rho \rho$ states.
The results for processes with the exchange of heavy mesons (compared to pion)
strongly depend on the details of the hadronic form factors.
By comparing the theoretical results and the cross sections
found in the CERN-ISR experiment \cite{Breakstone:1993ku}
we fixed the parameters of the off-shell meson form factor
and the $I\!\!P \sigma \sigma$ and $f_{2 I\!\!R} \sigma \sigma$ couplings.
In the case of $\sigma \sigma$ contribution we use 
two sets of the coupling constants;
standard (set A) and enhanced (set B) ones, 
see (2.11) and (2.12) of \cite{Lebiedowicz:2016zka}, respectively.
\footnote{There is quite a good agreement between our $\sigma \sigma$ result
with a monopole form factor and the $4 \pi$ ($J=0$, phase space) data
from \cite{Breakstone:1993ku}. Note that this implies that the set~B of couplings, 
which are larger than the corresponding pion couplings, seems to be preferred.}
In the case of $\rho \rho$ contribution 
the $\rho$ meson reggeization suppresses large masses of $M_{4 \pi}$ distributions.
This is also the case when the separation in rapidity 
between the two $\rho$ mesons increases, see Fig.~4 of \cite{Lebiedowicz:2016zka}.
\footnote{We have found that the diffractive mechanism
in $pp$ collisions considered by us
leads to the cross section for the $\rho \rho$ final state more than 
three orders of magnitude larger than the corresponding cross section
for $\gamma \gamma \to \rho \rho$ and double scattering photon-pomeron
(pomeron-photon) mechanisms considered in \cite{Goncalves:2016ybl}.}

\section{Conclusions}
\label{conclusions}

In our recent paper \cite{Lebiedowicz:2016ioh} we have analysed the exclusive central production of
dipion continuum and resonances contributing to the $\pi^{+} \pi^{-}$ pair production
in proton-(anti)proton collisions in an effective field-theoretic approach
with tensor pomerons and reggeons as proposed in \cite{Ewerz:2013kda}.
We have included the scalar ($f_{0}(500)$, $f_{0}(980)$) 
and tensor $f_{2}(1270)$ resonances
as well as the vector $\rho(770)$ resonance in a consistent way.
In the case of $f_{2}(1270)$-meson production via ``fusion'' of two tensor pomerons
we have found (see Appendix~A of \cite{Lebiedowicz:2016ioh}) 
the seven possible $I\!\!P I\!\!P f_{2}$ tensorial couplings.
The different couplings give different results due to 
different interference effects of the $f_{2}$ resonance 
and the dipion continuum contributions. 
We have shown that the resonance structures 
in the measured two-pion invariant mass spectra
depend on the cut on proton transverse momenta and/or 
on four-momentum transfer squared $t_{1,2}$ used in experiment.
The model parameters of the optimal $I\!\!P I\!\!P f_{2}$ coupling ($j=2$) have been roughly adjusted
to the recent CDF and preliminary STAR experimental data 
and then used for the predictions for the ALICE, and CMS experiments.
We have made estimates of cross sections for both the diffractive 
and photoproduction contributions.
We have shown some differential distributions related to produced pions
as well as some observables related to final state protons,
e.g., different dependence on proton transverse momenta and 
azimuthal angle correlations between outgoing protons
could be used to separate the photoproduction term, see \cite{Lebiedowicz:2016ioh}. 
The absorption effects due to $pp$ and $\pi p$ interactions, 
discussed in \cite{Lebiedowicz:2015eka},
lead to a significant modification of the shape of the distributions 
in $\phi_{pp}$, $p_{t,p}$, $t_{1,2}$ and it would therefore be useful 
to study such observables experimentally 
when measuring forward protons (STAR, ATLAS-ALFA, CMS-TOTEM).\\
To summarize: We have given a consistent treatment of 
the exclusive $\pi^{+}\pi^{-}$ and $\pi^{+}\pi^{-}\pi^{+}\pi^{-}$ production
in $pp$ collisions
in an effective field-theoretic approach.
A measurable cross section of order of a few $\mu b$ was obtained 
for both processes which should provide experimentalists
interesting challenges to check and explore it.

\subparagraph{Acknowledgement.}
This work was partially supported by the MNiSW Grant No.~IP2014~025173 (Iuventus Plus)
and the Polish National Science Centre grants DEC-2014/15/B/ST2/02528 and DEC-2015/17/D/ST2/03530.


\begin{thebibliography}{99}

\bibitem{Lebiedowicz:2016ioh}
P. Lebiedowicz, O. Nachtmann, A. Szczurek, Phys. Rev. D {\bf 93}, 054015 (2016).

\bibitem{Fiore:2015lnz}
R. Fiore, L. Jenkovszky, R. Schicker, Eur. Phys. J. C {\bf 76}, 38 (2016).

\bibitem{Austregesilo:2016sss}
A. Austregesilo (COMPASS Collaboration), AIP Conf. Proc. {\bf 1735}, 030012 (2016).


\bibitem{Breakstone:1986xd}
T.~{\AA}kesson \textit{et al.}, (AFS Collaboration), Nucl. Phys. B {\bf 264}, 154 (1986);
A.~Breakstone \textit{et al.} (ABCDHW Collaboration), Z. Phys. C {\bf 31}, 185 (1986);
A.~Breakstone \textit{et al.} (ABCDHW Collaboration), Z. Phys. C {\bf 42}, 387 (1989);
Erratum: Z. Phys. C {\bf 43}, 522 (1989);
A.~Breakstone \textit{et al.} (ABCDHW Collaboration), Z. Phys. C {\bf 48}, 569 (1990).

\bibitem{Adamczyk:2014ofa}
L. Adamczyk, W. Guryn, J. Turnau, Int. J. Mod. Phys. A {\bf 29}, 1446010 (2014).

\bibitem{Aaltonen:2015uva}
T. Aaltonen \textit{et al.} (CDF Collaboration), Phys. Rev. D {\bf 91}, 091101 (2015).

\bibitem{Schicker:2012nn}
R. Schicker (ALICE Collaboration), arXiv:1205.2588 [hep-ex].

\bibitem{CMS:2015diy}
CMS Collaboration, Report No. CMS-PAS-FSQ-12-004.

\bibitem{McNulty:2016sor}
R. McNulty, PoS(DIS2016)181, arXiv:1608.08103 [hep-ex].

\bibitem{Staszewski:2011bg}
R. Staszewski, P. Lebiedowicz, M. Trzebi{\'n}ski, J. Chwastowski, A. Szczurek,
Acta Phys. Polon. B {\bf 42}, 1861 (2011).

\bibitem{Szczurek:2009yk}
A. Szczurek and P. Lebiedowwicz, Nucl. Phys. A {\bf 826}, 101 (2009).

\bibitem{Lebiedowicz:2009pj}
P. Lebiedowicz and A. Szczurek, Phys. Rev. D {\bf 81}, 036003 (2010).

\bibitem{Lebiedowicz:2011nb}
P. Lebiedowicz, R. Pasechnik, A. Szczurek, Phys. Lett. B {\bf 701}, 434 (2011) .

\bibitem{Lebiedowicz:2011tp} 
P. Lebiedowicz and A. Szczurek, Phys. Rev. D {\bf 85}, 014026 (2012).

\bibitem{Lebiedowicz:2015eka}
P. Lebiedowicz and A. Szczurek, Phys. Rev. D {\bf 92}, 054001 (2015).

\bibitem{Ewerz:2013kda}
C. Ewerz, M. Maniatis, O. Nachtmann, Annals Phys. {\bf 342}, 31 (2014).

\bibitem{Nachtmann:1991ua}
O. Nachtmann, Annals Phys. {\bf 209}, 436 (1991).

\bibitem{Ewerz:2016onn}
C. Ewerz, P. Lebiedowicz, O. Nachtmann, A. Szczurek, arXiv:1606.08067 [hep-ph].

\bibitem{Adamczyk:2012kn}
L. Adamczyk \textit{et al.} (STAR Collaboration), Phys. Lett. B {\bf 719}, 62 (2013).

\bibitem{Lebiedowicz:2013ika}
P. Lebiedowicz, O. Nachtmann, A. Szczurek, Annals Phys. {\bf 344}, 301 (2014).

\bibitem{Bolz:2014mya}
A. Bolz, C. Ewerz, M. Maniatis, O. Nachtmann, M. Sauter, A. Sch{\"o}ning, JHEP {\bf 1501}, 151 (2015).

\bibitem{Lebiedowicz:2014bea}
P. Lebiedowicz, O. Nachtmann, A. Szczurek, Phys. Rev. D {\bf 91}, 074023 (2015).

\bibitem{Lebiedowicz:2016zka}
P. Lebiedowicz, O. Nachtmann, A. Szczurek, Phys. Rev. D {\bf 94}, 034017 (2016).

\bibitem{Breakstone:1993ku}
A. Breakstone \textit{et al.} (ABCDHW Collaboration), Z. Phys. C {\bf 58}, 251 (1993).

\bibitem{Goncalves:2016ybl}
V.P. Goncalves, B.D. Moreira, F.S. Navarra, Eur. Phys. J C {\bf 76}, 388 (2016).

\end{thebibliography}
\end{document}